\input epsf.tex
\newcommand\bfigt{\begin{figure}[top]}
\newcommand\efig{\end{figure}}
\newcommand{\ra}{\rightarrow}
\newcommand{\si}{\sigma}
\newcommand{\ta}{\tau}
\newcommand{\tn}{\otimes}
\newcommand{\ga}{\gamma}
\newcommand{\pl}{\partial}
\newcommand{\la}{\lambda}
\newcommand{\bq}{\begin{equation}}
\newcommand{\eq}{\end{equation}}
\newcommand{\ph}{\varphi}
\newcommand{\iy}{\infty}
\newcommand{\ps}{\psi}

\newcommand{\dl}{\delta}
\newcommand{\ch}{\raisebox{.4ex}{$\chi$}}
\newcommand{\ov}{\over}
\newcommand{\ve}{\varepsilon}

\newcommand{\rh}{\rho}
\newcommand{\ba}{\left(\begin{array}{cc}}
\newcommand{\ea}{\end{array}\right)}
\newcommand{\bdet}{\left|\begin{array}{cc}}
\newcommand{\edet}{\end{array}\right|}
\newcommand{\eph}{\varepsilon\varphi}
\newcommand{\eps}{\varepsilon\psi}
\newcommand{\be}{\beta}
\newcommand{\al}{\alpha}
\newcommand{\hf}{{1\ov2}}
\newcommand{\vet}{\tilde v_{\ve}}
\newcommand{\sep}{_{\ve}}

\newcommand{\cI}{{\cal I}}
\newcommand{\cR}{{\cal R}}

\newcommand{\cQ}{{\cal Q}}
\newcommand{\cP}{{\cal P}}

\newcommand{\norm}{\parallel}
\newcommand{\noi}{\noindent}
\newcommand{\Ltl}{\tilde{L_2}}
\newcommand{\Htl}{\tilde{H_1}}
\newcommand{\ds}{\oplus}
\newcommand{\iyy}{\int_{-\iy}^{\iy}}
\newcommand{\inv}{^{-1}}
\newcommand{\izy}{\int_0^{\iy}}
\newcommand{\lp}{\left(}
\newcommand{\rp}{\right)}
\newcommand{\bR}{{\bf R}}
\newcommand{\bch}{\raisebox{.4ex}{$\bar\chi$}}
\newcommand{\st}{_{\ta}}

\documentstyle[12pt]{article} \textwidth=6in \oddsidemargin=.3in
\begin{document}

\begin{center} {\large\bf On Orthogonal and Symplectic Matrix Ensembles}
\end{center}
\vspace{.1in}\begin{center}{{\bf Craig A.~Tracy}\\
{\it Department of Mathematics \\
        and\\
Institute of Theoretical Dynamics\\
University of California\\
Davis, CA 95616, USA\\
E-mail address: tracy@itd.ucdavis.edu}}\end{center}
\begin{center}{{\bf Harold Widom}\\
{\it Department of Mathematics\\
University of California\\
Santa Cruz, CA 95064, USA\\
E-mail address: widom@cats.ucsc.edu}}\end{center}
\vspace{.1in}
\begin{center}{\bf Abstract}\end{center}
The focus of this paper is on the probability, $E_\beta(0;J)$, that a set $J$
consisting
of a finite union of intervals contains no eigenvalues for  the finite $N$
Gaussian
Orthogonal  ($\beta=1$) and  Gaussian Symplectic  ($\beta=4$) Ensembles and
their
respective
scaling limits both in the bulk and at the edge of the
spectrum.  We show how these probabilities
can be expressed in terms of quantities arising in the corresponding unitary
($\beta=2$) ensembles. Our most explicit new results concern the distribution
of the
largest eigenvalue in each of these ensembles.  In the edge scaling limit we
show
that these largest eigenvalue distributions
are given in terms of a particular Painlev\'e II function.
\newpage
\begin{center}{\bf I. Introduction} \end{center}

In the standard random matrix
 models of $N\times N$ Hermitian or symmetric matrices the probability
density that the eigenvalues lie in infinitesimal intervals about the points
$x_1,\cdots,x_N$ is given by
\[P_\be(x_1,\cdots,x_N)=C_{N\be}\; e^{-\be\sum
V(x_i)}\,\prod_{j<k}|x_j-x_k|^{\be},\]
where the constant $C_{N\be}$ is such that the integral of the right side
equals $1$.
In the Gaussian ensembles the potential $V(x)$ equals $x^2/2$ and the cases
$\be=1,\,
2$ and $4$ correspond to the orthogonal, unitary, and symplectic ensembles,
respectively, since the underlying probability distributions
are invariant under these groups.

When $\be=2$ the polynomials orthogonal
with respesct to the weight function $e^{-2\,V(x)}$ play an important role. If
$\ph_i(x)\;(i=0,\,1\cdots)$ is the family of functions obtained by
orthonormalizing
the sequence $x^i\,e^{-V(x)}$, then
\[P_2(x_1,\cdots,x_N)=\det (K_N(x_i,x_j))\qquad(i,j=1,\cdots,N),\]
where
\bq K_N(x,y):=\sum_{i=0}^{N-1}\,\ph_i(x)\,\ph_i(y).\label{KN}\eq
It follows from this that the so-called ``$n$-level correlation function'' is
given by
\bq R_{n2}(x_1,\cdots,x_n)=\det
(K_N(x_i,x_j))\qquad(i,j=1,\cdots,n),\label{RN}\eq
and the ``$n$-level cluster function'' by
\bq T_{n2}(x_1,\cdots,x_n)=\sum K_N(x_{\si 1},x_{\si 2})\cdots
K_N(x_{\si(n-1)},x_{\si n})
\label{TN}\eq
where the sum is taken over all cyclic permutations $\si$ of the integers
$1,\cdots,n$, in some order. (See \cite{M},(5.1.2, 3) and (5.2.14, 15).)
The probability $E_2(0;J)$ that no eigenvalues lie in the set $J$ is equal to
the
Fredholm determinant of the integral operator on $J$ (more precisely, on
functions
on $J$) with kernel $K_N(x,y)$.
There are analogues of this for scaled limits of these ensembles. If one
takes a scaled limit in ``the bulk'' of the spectrum for the Gaussian unitary
ensemble
then (\ref{RN}) and (\ref{TN}) become
\bq R_{n2}(x_1,\cdots,x_n)=\det
(S(x_i,x_j))\qquad(i,j=1,\cdots,n),\label{R2}\eq
\bq T_{n2}(x_1,\cdots,x_n)=\sum S(x_{\si 1},x_{\si 2})\cdots
S(x_{\si(n-1)},x_{\si n}),
\label{T2}\eq
where
\[S(x,y):={1\ov\pi}{\sin(x-y)\ov x-y}.\]
Now $E_2(0,J)$ is the Fredholm determinant of the operator on $J$ with kernel
$S(x,y)$. If one scales the same ensemble at ``the edge'' of the spectrum this
is
replaced by the ``Airy kernel''
\[ { {\rm Ai}(x) {\rm Ai}'(y) - {\rm Ai}'(x) {\rm Ai}(y) \over
 x-y }\> ,\]
where ${\rm Ai}(x)$ is the Airy function, and if one scales the Laguerre
ensemble
(which corresponds to the choice of potential $V(x)=\hf x-\hf\al\log x$) at the
edge
one obtains the ``Bessel kernel''
\[{\ph(x)\ps(y)-\ps(x)\ph(y)\ov x-y},\qquad\ph(x)=J_{\al}(\sqrt x),\,
\ps(x)=x\,\ph'(x).\]

F.~J.~Dyson \cite{D} discovered that the introduction of so-called ``quaternion
determinants'' allows one to write down $\be=1$ and $\be=4$ analogues of
(\ref{R2})
and (\ref{T2}). We define $\ve(x):=\hf\, {\rm sgn}\,x$ and
\bq S(x):={\sin x\ov\pi x},\quad DS(x):=S'(x),\quad
IS(x):=\int_0^xS(y)\,dy,\quad JS(x):=IS(x)-\ve(x),\label{notation}\eq
\bq\si_1(x,y):=\ba S(x-y)&DS(x-y)\\&\\JS(x-y)&S(x-y)\ea,\label{si1}\eq
\bq\si_4(x,y):=\ba
S(2(x-y))&DS(2(x-y))\\&\\IS(2(x-y))&S(2(x-y))\ea.\label{si4}\eq
If these $2\times2$ matrices are thought of as quaternions then the analogues
of
(\ref{R2}) and (\ref{T2}) for $\be=1$ and $\be=4$ are obtained
by replacing $S(x,y)$ by $\si_{\be}(x,y)$, and by interpreting the right side
of
(\ref{R2}) as a quaternion determinant. The right side of (\ref{T2}) is a
scalar
quaternion, the scalar being equal to $\hf$ times the trace of the right side
when
it is interpreted as a matrix. It follows from these facts, by an argument to
be
found in A.7 of \cite{M}, that $E_\be(0;J)$ is equal to the square root of the
Fredholm determinant of the
integral operator on $J$ with matrix kernel $\si_{\be}(x,y)$. There are similar
but more complicated matrix kernels for the finite $N$ matrix ensembles
(\cite{M}, Chaps.~6--7).

The focus of this paper is $E_\be(0;J)$ for $\be=1$ and $4$ when $J$ is
a finite union of intervals, for both finite $N$ and scaled
Gaussian ensembles.
We shall show how these can all be expressed in terms of quantities arising in
the
corresponding  $\be=2$  ensembles. The $E_\be(0;J)$ are Fredholm determinants
of
certain matrix-valued kernels, and by manipulating these Fredholm
determinants we are able
to write them as ordinary scalar determinants whose order depends only on the
ensemble
and the number of intervals in $J$, and not on $N$ if the ensemble is finite.
The
entries of this determinant contain integrals involving the resolvent kernel
for the
$\be=2$ kernel (these are the ``quantities'' alluded to above). For the scaled
Gaussian
ensembles our  results for general $J$ are given below in (\ref{E1JDet}) and
(\ref{E4JDet}) and
for  the finite $N$ Gaussian ensembles the results for general $J$ are in
 ({\ref{1op}) and (\ref{4op}).

The evaluation of the
integrals appearing in our above quoted
expressions is  a separate matter. In the cases of greatest interest to
us---$J$
a finite interval for the scaled ensembles, $J$ a semi-infinite interval for
the
finite $N$ ensembles---there are systems of differential equations associated
with
those integrals. The equations are easily solved in the cases of the scaled
ensembles
and we recover known formulas for the probability in these models of the
absence of
eigenvalues in an interval. (These are found in
\cite{M} as formulas (6.5.19) and (10.7.5).)  It then follows from \cite{JMMS}
that
all these probabilities  are expressible in terms of a
Painlev\'e function of fifth kind (see also \cite{BTW,Intro}).  For finite
$N$, and $J$ the semi-infinite interval $(t,\,\iy)$, $E_\be(0;J)$ is the
probability
distribution function $F_{N\be}(t)$ for the largest eigenvalue.
In \cite{FD} we showed that when $\be=2$ this is expressible in terms of a
Painlev\'e
function (this time $P_{IV}$) and we hoped to be able to find representations
in the cases
$\be=1$ and $\be=4$ also, but we were unable to solve the associated system of
differential
equations. However we succeeded for their limits scaled at ``the edge ''
because the
equations scale to a system which we can solve. To explain our results, we
recall
that  $F_{N\be}(2\sigma\sqrt{N}+t)$
tends to the Heaviside function as $N\ra\iy$
where $\sigma$ is the standard deviation of the Gaussian
distribution on the off-diagonal matrix elements \cite{BY}.
(Our choice of $P_\be(x_1,\cdots,x_N)$
corresponds to a standard deviation $\sigma=1/\sqrt{2}$ which
is the usual choice \cite{M}.  We also recall that this
result holds for the larger class of  so-called Wigner matrices \cite{BY}, but
the
results that follow are only known for the Gaussian ensembles.)\ \
This says, roughly, that the largest eigenvalue is within $o(1)$ of
$2\sigma\sqrt{N}$, and so
this is thought of as the right edge of the spectrum. In fact, the largest
eigenvalue
is within $O(N^{-1/6})$ of $2\sigma\sqrt{N}$ and we consider here the more
refined limits
\[F_{\be}(s):=\lim_{N\ra\iy}F_{N\be}(2\sigma\sqrt{N}+{\sigma s\ov N^{1/6}}).\]
(As the notation suggests,  $F_\beta$ is independent
of $\sigma$.)\ \
In earlier work \cite{Airy}  we showed that when $\be=2$
this is given in terms of
another Painlev\'e function ($P_{II}$---see (\ref{2pdf}) below).
Now we shall find representations for
$F_1(s)$ and $F_4(s)$ in terms of this same function (see (\ref{1pdf}) and
(\ref{4pdf}) below).
The probability densities
$f_\be(s) = dF_\be/ ds $
are graphed in Fig.~1.  Though these results strictly apply only in the limit
as the size
of the matrices tends to infinity,  simulations of finite $N$ GOE, GUE and GSE
matrices show that the empirical probability density of the largest eigenvalue
is
well approximated by $f_\be$, $\be=1,2,4$, for $N\ge 200$.
\bfigt\vspace*{-50mm}\hspace*{15mm}\epsfysize=180mm \epsffile{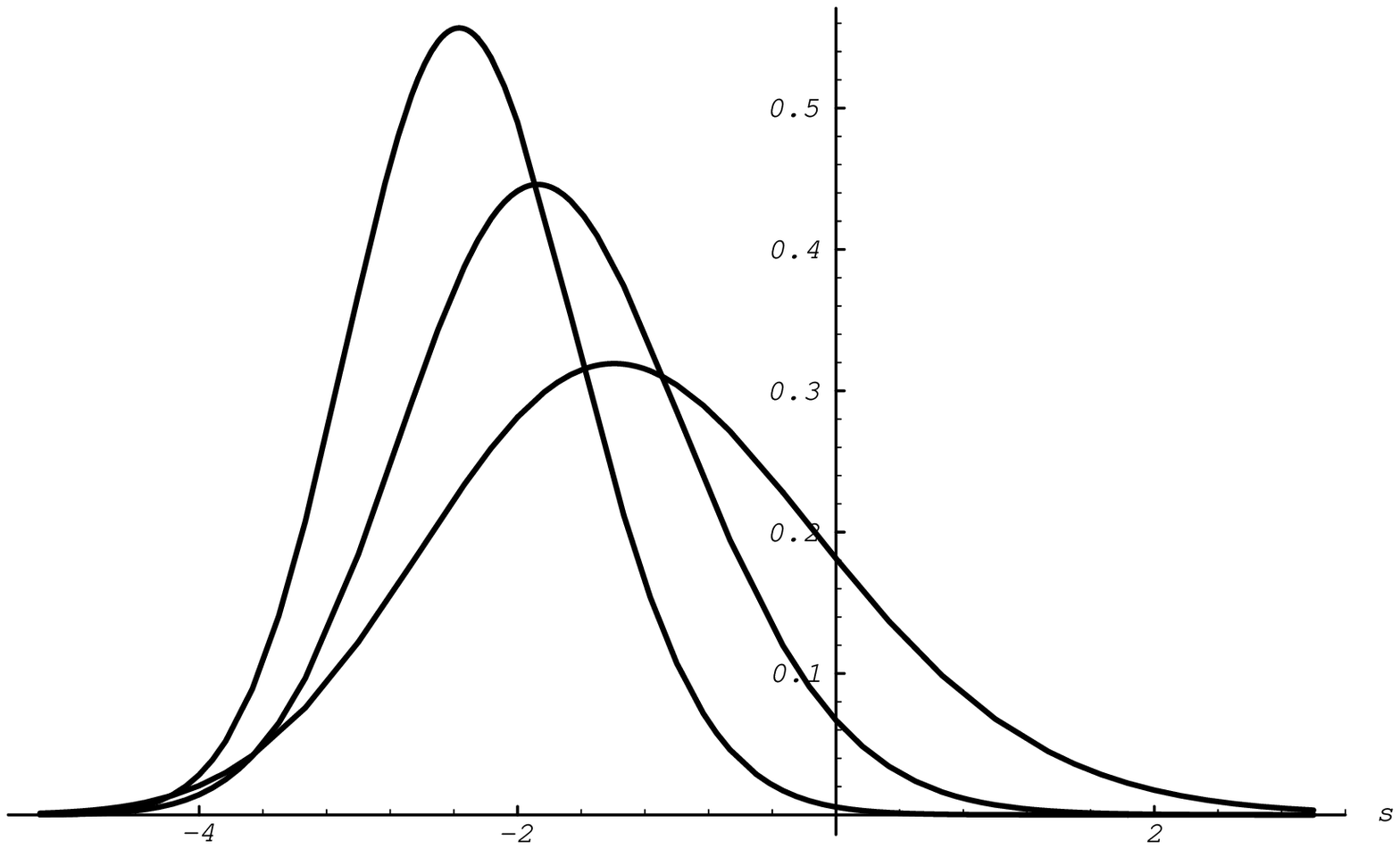}
\vspace*{-50mm}
\caption{The probability densitities $f_\beta(s)=
{dF_\beta/ ds}$, $\beta=1$, $2$,
and $4$ for the position of the largest eigenvalue in the edge scaling limit.
The probability
density with the smaller value of $\beta$ (``higher temperature'') has the
larger variance.}
\efig

Although this paper treats exclusively the Gaussian ensembles, the methods
appear
quite general and should apply to other ensembles as well. In particular one
might
expect to be able to express the limiting distribution function for the
smallest
eigenvalue in the $\be=1$ and 4 Laguerre ensembles in terms of a $P_{III}$
function,
as is the case for $\be=2$ \cite{Bessel}.

In Sections II and III we derive our expressions for $E_\be(0;J)$ for the
bulk-scaled
Gaussian ensembles when $J$ is a finite union of intervals. In
Section IV we specialize to the case of one interval and see how to recover the
formulas cited above for the probability of absence of eigenvalues.
Section V contains the analogous derivations, for general $J$, for the finite
$N$
ensembles. These are more complicated than in the scaled case, because the
expressions for the analogous matrix functions $\si_{N\be}(x,y)$ have ``extra''
terms.
In Section VI we derive the differential equations associated with these
ensembles
when $J$ is a semi-infinite interval, and in Section VII we derive
the results on the limiting probability distribution for the largest
eigenvalues in
the orthogonal and symplectic ensembles.

To obtain our formulas for the Fredholm determinants we think of the operators
with
matrix kernel instead as
matrices with operator kernels, and then manipulate their determinants
($2\times2$
determinants with operator entries) in a way the reader might find suspect.
These
manipulations are, however, quite correct and in the last section we present in
detail
their justification for the bulk-scaled ensembles. For the basic definitions
and properties of operator determinants we refer the reader to
Chapter IV of \cite{GK},
where everything we use will be found.

There is an alternative route to the results of Section VII, based on the fact
that the limiting probability distributions are Fredholm determinants involving
the
scaled kernel, the Airy kernel.  We did not choose this route because we would
have had
to give yet another derivation of a set of differential equations for the
entries of
a scalar determinant, and because we would have had to present yet another
justification
for the manipulation of the Fredholm determinants, in this case involving the
Airy kernel.
This would be more delicate than the justification in Section VIII for the sine
kernel
because the Airy kernel is not as nicely behaved at $-\infty$.  (We do  use the
Airy kernel in our derivation, but only on intervals semi-infinite on the
right, where
it is very well behaved.)

\begin{center}{\bf II. The Scaled Gaussian Orthogonal Ensemble}
\end{center}\par

First,  we introduce some terminology.
If $K(x,y)$ is the kernel of an integral operator $K$
then we shall speak interchangeably of the determinant {\em for\/} $K(x,y)$ or
$K$, and
the determinant {\em of\/} the operator $I-K$.
We denote by $S$ and $\ve$ the integral operators on the entire
real line {\bf R} with kernels $S(x-y)$ and $\ve(x-y)$,
respectively (see (\ref{notation})), and write $D$ for $d/dx$. Notice that
$DS$ has kernel $DS(x-y)$ (fortunately!) and it follows from the evenness of
$S(x)$
that $IS(x-y)$ and $JS(x-y)$ are the kernels of the operators $\ve S$ and
$\ve S-\ve$, respectively. We denote by $\ch$ the operator of
multiplication by $\ch_J(x)$, the characteristic function of $J$.

$E_1(0;J)^2$ equals the determinant for the kernel $\si_1(x,y)$ on
$J$, or, what is the same thing, the determinant for the operator with kernel
\[\ch_J(x)\,\ba S(x-y)&DS(x-y)\\&\\JS(x-y)&S(x-y)\ea\,\ch_J(y)\]
on {\bf R}. This can be represented as the $2\times2$ operator matrix
\bq\ba \ch S\ch&\ch DS\ch\\&\\\ch(\ve S-\ve)\ch&\ch S\ch\ea.\label{1Op}\eq
Since $D\ve=I$, the identity operator, this can be factored as
\bq\ba \ch D&0\\&\\0&\ch\ea\ba \ve S\ch&S\ch\\&\\(\ve
S-\ve)\ch&S\ch\ea.\label{1fac}\eq

It is a general fact that for operators $A$ and $B$ the determinants for $AB$
and
$BA$ are equal. So we can take the factor on the left above and bring it around
to the
right, combine the two factors $\ch$ into one, and find that the above can be
replaced by
\bq\ba \ve S\ch D&S\ch\\&\\(\ve S-\ve)\ch D&S\ch\ea.\label{1repl}\eq
Subtracting row 1 from row 2 and then adding column 2 to column 1 we see that
the
determinant for this is the same as that for
\bq\ba\ve S\ch D+S\ch&S\ch\\&\\-\ve\ch D&0\ea,\label{1Red}\eq
and so equals the determinant of the operator
\[\ba I-\ve S\ch D-S\ch&-S\ch\\&\\\ve\ch D&I\ea.\]
Next we subtract column 2, right-multiplied by $\ve\ch D$, from column 1 and
then
add row 2, left-multiplied by $S\ch$, to row 1. (Column operations are always
associated with right-multiplication and row operations with left.) The result
is
\[\ba I-\ve S\ch D-S\ch+S\ch\ve\ch D&0\\&\\0&I\ea,\]
and the determinant of this equals
\bq\det\,(I-\ve S\ch D-S\ch+S\ch\ve\ch D).\label{1Det}\eq

So we have shown that $E_1(0;J)^2$ equals the determinant of the operator with
scalar kernel
\[I-\ve S\ch D-S\ch+S\ch\ve\ch D=I-S\ch-S(I-\ch)\ve\ch D,\]
where we used here the fact that $\ve$ and $S$ commute.
Now $I-S\ch$ is precisely the operator which arises in the bulk-scaled Gaussian
unitary ensemble. (In particular its determinant, which is the same as the
determinant
of $I-\ch S\ch$, is exactly $E_2(0;J)$.)\ \ Because $J$ is a finite union of
intervals
the last summand will turn out to be a finite rank operator, so if we factor
out
$I-S\ch$, whose determinant we know, then we obtain an operator of the form
$I-$ finite rank operator, whose determinant is just a
numerical determinant. Of course this factoring out requires introduction of
the
inverse $(I-S\ch)^{-1}$  which is why the resolvent kernel for $S\ch$ appears
in the
final result.

Now for the details. We denote by $R$ the resolvent operator for $S\ch$, so
that
\[(I-S\ch)^{-1}=I+R,\]
and by $R(x,y)$ its kernel. Observe that this is smooth in $x$ but
discontinuous
in $y$. Factoring out $I-S\ch$ and using the fact that
determinants multiply, we obtain
\bq E_1(0;J)^2=E_2(0;J)\;\det\,(I-(S+RS)(1-\ch)\ve\ch D).\label{E1J}\eq
To write the last operator more explicitly we denote the intervals comprising
$J$ by
$(a_{2k-1},a_{2k}),\;(k=1,\cdots,m)$, and set
\[\ve_k(x):=\ve(x-a_k),\quad\dl_k(y):=\dl(y-a_k),\quad R_k(x):=R(x,a_k),\]
where the last really means
\[\lim_{\stackrel{y\ra a_{k}}{y\in J}}R(x,y).\]
We use the notation $\al\tn\be$ for the operator with kernel $\al(x)\,\be(y)$,
the
most general finite rank operator being a sum of these. For any operator $A$ we
have
\bq A(\al\tn\be)=(A\al\tn\be),\quad (\al\tn\be)A=(\al\tn A^t\be),\label{ten}\eq
where $A^t$ is the transpose of $A$.

It is easy to see that the commutator $[\ch,D]$ has the representation
\[[\ch,D]=\sum_{k=1}^{2m}\,(-1)^k\dl_k\tn\dl_k,\]
and so
\bq\ve[\ch,D]=\sum_{k=1}^{2m}\,(-1)^k\ve_k\tn\dl_k.\label{epcom}\eq
Since $\ve D=I$ we have
\[(1-\ch)\ve\ch D=(1-\ch)\ve[\ch,D],\]
and we deduce the representation
\[(S+RS)(1-\ch)\ve\ch D=\sum_{k=1}^{2m}(-1)^k(S+RS)(1-\ch)\ve_k\tn\dl_k.\]

The determinant for this is evaluated using the general formula
\bq\det\,(I-\sum_{k=1}^n\al_k\tn\be_k)=
\det\,\Bigl(\dl_{j,k}-(\al_j,\be_k)\Bigr)_
{j,k=1,\cdots,n},\label{detform}\eq
where $(\al_j,\be_k)$ denotes the inner product. (This is an exercise in linear
algebra. Its generalization to an arbitrary trace class operator is in
\cite{GK}.)\ \
Those that arise in our case are
\bq((S+RS)(1-\ch)\ve_j,\dl_k)=((1-\ch)\ve_j,(S+S\,R^t)
\dl_k)=((1-\ch)\ve_j,R_k),
\label{SRS}\eq
where we have used the fact that $(S+S\,R^t)\ch=R$. (This is perhaps most
quickly seen
by writing $R=\sum_{k=1}^{\iy}(S\ch)^k$ and using $S=S^t$.)\ \ Thus we have
established
the formula
\bq E_1(0;J)^2=E_2(0;J)\;\det\,\Bigl(\dl_{j,k}-(-1)^k((1-\ch)\ve_j,R_k)\Bigr)_
{j,k=1,\cdots,2m}.\label{E1JDet}\eq

\begin{center}{\bf III. The Scaled Gaussian Symplectic Ensemble} \end{center}

Because of the factor 2 in the arguments of the functions in the matrix
(\ref{si4})
we make the variable changes $x\ra x/2,\,y\ra y/2$, and find that $E_4(0;J/2)$
is the
square root of the determinant for operator with kernel
\[\hf\ch_J(x)\,\ba S(x-y)&DS(x-y)\\&\\IS(x-y)&S(x-y)\ea\,\ch_J(y).\]
Thus the operator matrix (\ref{1Op}) of the last section is replaced by
\[\hf\ba \ch S\ch&\ch DS\ch\\&\\\ch\ve S\ch&\ch S\ch\ea.\]
Proceeding exactly as before we find that (\ref{1Red}) is replaced by
\[\hf\ba\ve S\ch D+S\ch&S\ch\\&\\0&0\ea\]
and (\ref{1Det}) by
\[\det\,(I-\hf S\ve\ch D-\hf S\ch).\]
But notice that the operator here equals
\[I-S\ch-\hf S\ve[\ch,D].\]
Using this we factor out $I-S\ch$ as before and conclude that the analogue of
(\ref{E1J}) is
\bq E_4(0;J/2)^2=E_2(0;J)\;\det\,(I-\hf(I+R)S\ve[\ch, D]).\label{4prod}\eq
Using (\ref{epcom}) once again we see that
\[(S+RS)\ve [\ch,D] =\sum_{k=1}^{2m}(-1)^k(S+RS)\ve_k\tn\dl_k,\]
and we obtain the analogue of (\ref{E1JDet}),
\bq E_4(0;J/2)^2=E_2(0;J)\;\det\,\Bigl(\dl_{j,k}-{1\ov
2}(-1)^k(\ve_j,R_k)\Bigr)_
{j,k=1,\cdots,2m} \label{E4JDet}\eq
where we used the fact that in the inner product we may use the identity
$\delta_k= \chi \delta_k$ since all evaluations are done by taking the
limit from within the interval.

\begin{center}{\bf IV. The Case} \boldmath $J=(-t,t)$\end{center}\par
In this case we have $m=1,\,a_1=-t,\,a_2=t$, and
\[\ve_1(x)=\ve(x+t),\;\ve_2(x)=\ve(x-t),\;R_1(x)=R(x,-t),\;R_2(x)=R(x,t).\]
If we define
\[\cI_{\pm}:=\int_{-\iy}^\iy R(x,t)\,\ve(x\mp t)\,dx\]
then
\bq-(\ve_1,R_1)=(\ve_2,R_2)=\cI_+,\quad-(\ve_2,R_1)=
(\ve_1,R_2)=\cI_-.\label{Iep}\eq
The first equalities above used the evenness of $R$ (i.e., the fact
$R(-x,y)=R(x,-y)$)
and the oddness of $\ve$.

To evaluate these integrals we find expressions for their derivatives with
respect
to $t$. Observe that $R(x,y)$ depends on the interval $J$ as well as $x$ and
$y$.
Formulas (2.9) and (2.22) of \cite{FD} give, for $y\in J$,
(see also Lemma 2 in \cite{Intro})
\[{\pl\ov\pl t}R(x,y)=R(x,-t)\,R(-t,y)+R(x,t)\,R(t,y),\]
\[\left({\pl\ov\pl x}+{\pl\ov\pl
y}\right)R(x,y)=R(x,-t)\,R(-t,y)-R(x,t)\,R(t,y),\]
from which it follows that
\[{\pl\ov\pl t}R(x,t)=2\,R(x,-t)\,R(-t,t)-{\pl\ov\pl x}R(x,t).\]
Hence, applying the product formula to the integrand and using
$\pl\ve(x\mp t)/\pl t=\mp\dl(x\mp t)$, we find that
\[{d\ov dt}\cI_{\pm}=\int_{-\iy}^\iy \,[2\,R(x,-t)\,R(-t,t)-{\pl\ov\pl
x}R(x,t)]\,
\ve(x\mp t)\,dx\mp R(\pm t,t).\]
Integrating by parts the term involving $\pl R(x,t)/\pl x$ and using parity
give
\[{d\ov dt}\cI_{\pm}=-2\,R(-t,t)\,\cI_{\mp}+R(\pm t,t)\mp R(\pm t,t).\]
Adding these two identities gives for $\cI_++\cI_-$ the differential equation
\[{d\ov dt}(\cI_++\cI_-)=-2\,R(-t,t)\,(\cI_++\cI_-)+2\,R(-t,t),\]
whose general solution is
\[1+c\,e^{-2\int_0^tR(-\ta,\ta)\,d\ta}.\]
(The function $R$ in the integral is the resolvent kernel for the interval
$(-\ta,\ta)$.)\ \ Since $\cI_{\pm}(0)=0$ we have $c=-1$ and so
\bq\cI_++\cI_-=1-\,e^{-2\int_0^tR(-\ta,\ta)\,d\ta}.\label{Isum}\eq
Similarly if we subtract the two identities for $d\cI_{\pm}/dt$ we obtain
\[{d\ov dt}(\cI_+-\cI_-)=2\,R(-t,t)\,(\cI_+-\cI_-)-2\,R(-t,t),\]
whose solution is
\bq\cI_+-\cI_-=1-\,e^{2\int_0^tR(-\ta,\ta)\,d\ta}.\label{Idiff}\eq

Because of (\ref{Iep}) these relations determine the inner products
$(\ve_j,R_k)$ which
arise in (\ref{E4JDet}). For the determinant in (\ref{E1JDet})
we need the inner products $((1-\ch)\ve_j,R_k)$. But observe that
\[((1-\ch)\ve_1,R_2)=((1-\ch)\ve_2,R_2)=\hf\int_{J^c}\,R(x,t)\,\mbox{sgn}\
x\,dx
={\cI_++\cI_-\ov2}={1-\,e^{-2\int_0^tR(-\ta,\ta)\,d\ta}\ov2},\]
\[((1-\ch)\ve_1,R_1)=((1-\ch)\ve_2,R_1)=-{\cI_++\cI_-\ov2}
={e^{-2\int_0^tR(-\ta,\ta)\,d\ta}-1\ov2}.\]
So we have all the inner products we need.

It follows from the last displayed formulas that the determinant
on the right side of (\ref{E1JDet}) equals
\bq\bdet{1+\,e^{-2\int_0^tR(-\ta,\ta)\,d\ta}\ov2}&
{1-\,e^{-2\int_0^tR(-\ta,\ta)\,d\ta}\ov2}
\\&\\{1-\,e^{-2\int_0^tR(-\ta,\ta)\,d\ta}\ov2}&
{1+\,e^{-2\int_0^tR(-\ta,\ta)\,d\ta}\ov2}\edet=e^{-2\int_0^tR(-\ta,\ta)\,d\ta}.
\label{1ratio}\eq
Now in this case $J=(-t,t)$ we have
\[{d\ov dt}\log\,E_2(0;J)=-2R(t,t),\]
and so
\bq E_2(0;J)=e^{-2\int_0^tR(\ta,\ta)\,d\ta}.\label{E2rep}\eq
Thus from (\ref{E1JDet}) and this fact we deduce
\bq E_1(0;J)=e^{-\int_0^t[R(\ta,\ta)+R(-\ta,\ta)]\,d\ta}.\label{Dpm}\eq

For the symplectic ensemble in this special case we use (\ref{Iep}),
(\ref{Isum}) and
(\ref{Idiff}) to evaluate the inner products appearing in the determinant on
the right
side of (\ref{E4JDet}) and find that it equals
\[\bdet 1-{\cI_+\ov2}&{\cI_-\ov2}\\&\\{\cI_-\ov2}&1-{\cI_+\ov2}\edet
=\left(1-{\cI_+\ov2}\right)^2-\left({\cI_-\ov2}\right)^2\]
\bq=\left(1-{\cI_++\cI_-\ov2}\right)\,\left(1-{\cI_+-\cI_-\ov2}\right)
=\left({e^{-\int_0^tR(-\ta,\ta)\,d\ta}+
e^{\int_0^tR(-\ta,\ta)\,d\ta}\ov2}\right)^2,
\label{4ratio}\eq
by (\ref{Isum}) and (\ref{Idiff}).
Hence, from (\ref{E4JDet}) and (\ref{E2rep}),
\bq E_4(0;J/2)=\hf\left(e^{-\int_0^t[R(\ta,\ta)+R(-\ta,\ta)]\,d\ta}+
e^{-\int_0^t[R(\ta,\ta)-R(-\ta,\ta)]\,d\ta}\right).\label{Dp/m}\eq

To see that (\ref{Dpm}) and (\ref{Dp/m}) are just the known formulas (6.5.19)
and
(10.7.5) of \cite{M} we recall the notations
\[D(t):=\prod_{i=0}^{\iy}(1-\la_i),\quad
D_+(t):=\prod_{i=0}^{\iy}(1-\la_{2i}),\quad
D_-(t):=\prod_{i=0}^{\iy}(1-\la_{2i+1}),\]
where $\la_0>\la_1>\cdots$ are the eigenvalues of the kernel $S(x-y)$ on
$(-t,t)$. Of
course $D(t)$ is the determinant for $S(x-y)$ while $D_{\pm}(t)$ are the
determinants
for the kernels
\[S_{\pm}(x,y):=\hf(S(x-y)\pm S(x+y)).\]
The resolvent kernels for these are
\[R_{\pm}(x,y)=\hf(R(x,y)\pm R(x,-y))\]
and from these we obtain the integral representations
\bq D_{\pm}(t)=e^{-\int_0^t[R(\ta,\ta)\pm R(-\ta,\ta)]\,d\ta}\label{Dpmrep}\eq
analogous to (\ref{E2rep}). Hence (\ref{Dpm}) and (\ref{Dp/m}) are (6.5.19) and
(10.7.5), respectively, of \cite{M}.

We mention here that the left side of (\ref{Idiff}) equals
\[-\int_J\,R(x,t)\,dx\]
and so in view of (\ref{Dpmrep}) the formula can be rewritten
\[{D_-(t)\ov D_+(t)}=1+\int_J\,R(x,t)\,dx.\]
This is equivalent to identity (A.16.6) of \cite{M} and, although this might
not be
obvious on first comparing the two, our derivation of it has
elements in common with the derivation, attributed to M.~Gaudin,
given in \cite{M}.

Finally, we mention that (\ref{E2rep}) and (\ref{Dpmrep})
may be expressed  in terms of Painlev\'e V following
\cite{JMMS}. (See  \cite{Intro} for a simplified treatment.)

\begin{center} {\bf V. The Finite \boldmath $N$ Gaussian Ensembles}
\end{center}

The analogue of the matrix kernel (\ref{si1}) for the finite $N$ GOE is given
by (6.3.8)
of \cite{M} and so we can write down the analogue of (\ref{1Op}). We shall
denote here
by $S$ the operator with kernel $K_N(x,y)$ given by
(\ref{KN}). Recall that the determinant for this operator on $J$ equals
$E_2(0;J)$
for the finite $N$ GUE. We also write
\[\ph(x)=({N\ov2})^{1/4}\,\ph_N(x),\quad\ps(x)=({N\ov2})^{1/4}\,\ph_{N-1}(x).\]
Then for $N$ even (this case is slightly simpler) the analogue of (\ref{1Op})
is
\bq\ch\ba S+\ps\tn\ve\ph&SD-\ps\tn\ph\\&\\\ve S-\ve+\ve\ps\tn\ve\ph
&S+\ve\ph\tn\ps\ea\ch.\label{1NOp}\eq
(See \cite{M}, Sec. 6.3.)
Now we have to be careful because $S$ does not commute with $\ve$ and $D$. But
we have
the simple relations
\bq[S,\,D]=\ph\tn\ps+\ps\tn\ph,\quad[\ve,\,S]=
-\eph\tn\eps-\eps\tn\eph.\label{coms}\eq
The first follows from (2.20) of \cite{FD} (where finite $N$ GUE is called the
``Hermite ensemble''), and the second follows
from the first upon left- and right-multiplying by $\ve$. (We used (\ref{ten})
here
and the antisymmetry of $\ve$. We shall make similar use of this below.)\ \
We write the first row of the matrix in (\ref{1NOp}) as
\[D\,(\ve S+\eps\tn\eph,\;\ve SD-\eps\tn\ph)
=D\,(S\ve-\eph\tn\eps,\; S+\eph\tn\ps),\]
where we applied (\ref{coms}). Using this, and applying
(\ref{coms}) now to the lower left corner, we find
that the analogue  of the second matrix in (\ref{1fac}) is
\[\ba(S\ve-\eph\tn\eps)\ch&(S+\eph\tn\ps)\ch\\&\\(S\ve-\ve-\eph\tn\eps)\ch
&(S+\eph\tn\ps)\ch\ea.\]
Now we  move that matrix $\ba \ch D&0\\0&\ch\ea$ around to the right and
find the analogue of the right side of (\ref{1repl}),
\[\ba (S\ve-\eph\tn\eps)\ch D&(S+\eph\tn\ps)\ch\\&\\
(S\ve-\ve-\eph\tn\eps)\ch D&(S+\eph\tn\ps)\ch\ea.\]
The same row and column operations as before reduce this to
\[\ba(S\ve-\eph\tn\eps)\ch D+(S+\eph\tn\ps)\ch&(S+\eph\tn\ps)\ch\\&\\
-\ve\ch D&0\ea,\]
the analogue of (\ref{1Red}). We complete the computation just as before and
find that
the determinant for this equals the determinant of the scalar operator
\[I-S\ve\ch D+(\eph\tn\eps)\ch D+(S+\eph\tn\ps)\ch(\ve\ch D-I)\]
\[=I-S\ch-S(1-\ch)\ve\ch D-\eph\tn\ch\ps
-\left(\eph\tn\ps\right)\,(1-\ch)\ve\ch D.\]
Factoring out $I-S\ch$ shows that the determinant of the above equals
$E_2(0,J)$
times the determinant of
\[I-(S+RS)(1-\ch)\ve\ch D-Q\sep\tn\ch\ps-Q\sep\tn\ps\,(1-\ch)\ve\ch D,\]
where \[Q\sep:=(I-S\ch)^{-1}\eph\] and we have used the same notation $R$ as
before
for the resolvent operator for $S\ch$. Using (\ref{epcom}) and the general fact
$(\al\tn\be)\,(\ga\tn\dl)=(\be,\ga)\,\al\tn\dl$ we see that the above operator
equals
\bq I-\sum_{k=1}^{2m}(-1)^k(S+RS)(1-\ch)\ve_k\tn\dl_k-Q\sep\tn\ch\ps
-\sum_{k=1}^{2m}(-1)^k\,(\ps,(1-\ch)\ve_k)\,Q\sep\tn\dl_k.\label{1op}\eq
Thus $E_1(0;J)^2/E_2(0;J)$ equals the determinant of this operator.

The analogue of the matrix kernel (\ref{si4}) for finite $N$ GSE is given
by (7.1.5) of \cite{M} where now  $N$ must be  odd.\footnote{In finite $N=2n+1$
GSE
the matrices are $2n\times 2n$ Hermitian matrices with each eigenvalue doubly
degenerate
\cite{M}.}
Because of the factor $\sqrt{2}$ in the arguments
of the functions in the matrix (7.1.5) of \cite{M}, we make the change
of variables $x\ra x/\sqrt{2}$, $y\ra y/\sqrt{2}$ and find that
$E_4(0;J/\sqrt{2})$
is the square root of the determinant for
\[ {1\ov 2} \ch\ba S+\ps\tn\ve\ph&SD-\ps\tn\ph\\&\\\ve S+\ve\ps\tn\ve\ph
&S+\ve\ph\tn\ps\ea\ch.\]
Proceedingly analogously to finite $N$ GOE
leads to the following formula for the operator whose
determinant is $E_4(0;J/\sqrt{2})^2/E_2(0;J)$:
\bq I-\hf\sum_{k=1}^{2m}(-1)^k(S+RS)\ve_k\tn\dl_k-Q\sep\tn\ch\ps
-\hf\sum_{k=1}^{2m}(-1)^k\,(\ps,\ve_k)\,Q\sep\tn\dl_k.\label{4op}\eq

\begin{center} {\bf VI. The Case} \boldmath $J=(t,\iy)$\end{center}

Now $J$ has the end-point $a_1=t$ and $a_2=\iy$, and we write
$\dl_t,\,\dl_{\iy},\,
\ve_t,\,\ve_{\iy},\,R_t$ and $R_{\iy}$ for the quantities $\dl_k,\,\ve_k,\,R_k$
($k=1,\,2$)
of the last sections. Note that
\[\ve_{\iy}=-\hf,\quad(1-\ch)\,\ve_t=(1-\ch)\,\ve_{\iy}=-\hf\,(1-\ch),\quad
R_{\iy}=0.\]
With this notation the operators (\ref{1op}) and (\ref{4op}) become
\bq
I-Q\sep\tn\ch\ps-\hf[(S+RS)(1-\ch)+(\ps,(1-\ch))\,Q\sep]\tn(\dl_t-\dl_{\iy}),
\label{1ops}\eq
\bq I-Q\sep\tn\ch\ps+\hf[(S+RS)\ve_t+(\ps,\ve_t)\,Q\sep]\tn\dl_t
+{1\ov 4}[(S+RS)1+(\ps,\,1)\,Q\sep]\tn\dl_{\iy},\label{4ops}\eq
respectively. Both operators are of the form
\bq I-\sum_{k=1}^n\al_k\tn\be_k\label{tenrep}\eq
so to evaluate their determinants using (\ref{detform}) we have many inner
products
to evaluate. We shall introduce several new quantities now and express all the
inner products, and therefore the determinants, in terms of them. Then we shall
write
down systems of linear differential equations (in the variable $t$) which in
principle
determine these quantities.

First, there are
\[Q:=(I-S\ch)^{-1}\ph,\quad P:=(I-S\ch)^{-1}\ps,\quad
Q\sep:=(I-S\ch)^{-1}\eph,\quad P\sep:=(I-S\ch)^{-1}\eps,\]
the third of which we have already met. We use small letters to denote the
values
of these functions at $x=t$:
\bq q=Q(t),\quad p=P(t),\quad q\sep=Q\sep(t),\quad
p\sep=P\sep(t).\label{qetc}\eq
(The functions $q$ and $p$ play important roles in the investigation of
$E_2(0;J)$
\cite{FD} and we think of them here as known.)\ \

Next, there are the inner products
\[u\sep:=(Q,\ch\eph)=(Q\sep,\ch\ph),\quad v\sep:=(Q,\ch\eps)=(P\sep,\ch\ps),\]
\[\vet:=(P,\ch\eph )=(Q\sep,\ch\ps ),\quad w\sep:=(P,\ch\eps )=(P\sep,\ch\ps
),\]
the last four being analogous to (2.4)--(2.5) of \cite{FD} with $x^j$ replaced
by
$\ve$. Our first system of differential equations (in which $q$ and $p$ appear
as
coefficients) will connect these with $q\sep$ and $p\sep$.

Finally there are two triples of integrals
\[\cR_1:=\int_{-\iy}^tR(x,t)\,dx,\quad\cP_1:=\int_{-\iy}^t
P(x)\,dx,\quad\cQ_1:=
\int_{-\iy}^t Q(x)\,dx,\]
\[\cR_4:=\iyy\ve_t(x)\,R(x,t)\,dx,\quad\cP_4:=
\iyy\ve_t(x)\,P(x)\,dx,\quad\cQ_4:=
\iyy\ve_t(x)\,Q(x)\,dx.\]
(The subscripts 1 and 4 indicate that these arise in GOE and GSE,
respectively.)
We shall find systems of differential equations for each of these triples.

\medskip\noi {\it The determinants of (\ref{1ops}) and (\ref{4ops}):}\medskip

We consider the GOE operator (\ref{1ops}) first. If we set
\[a_1:=(\ps,\,1-\ch)\]
then the operator is of the form (\ref{tenrep}) with $n=2$ and
\[\al_1=Q\sep,\quad \al_2=\hf[(S+RS)(1-\ch)+a_1\,Q\sep],\quad
\be_1=\ch\,\ps,\quad \be_2=\dl_t-\dl_{\iy}.\]
We also set
\bq c_{\ph}:=\ve\ph(\iy)=\hf\iyy\ph(x)\,dx,\quad
c_{\ps}:=\ve\ps(\iy)=\hf\iyy\ps(x)
\,dx.\label{cdef}\eq
For $N$ even a computation gives
\[ c_\ph= (\pi N)^{1/4} 2^{-3/4-N/2}\;{(N!)^{1/2}\ov (N/2)!}\; . \]
Since $Q\sep=\eph+S\ch(I-S\ch)^{-1}\eph$, we have $Q\sep(\iy)=c_{\ph}$.
Therefore,
with the notations introduced above,
\[(\al_1,\,\be_1)=\vet,\quad (\al_1,\,\be_2)=q\sep-c_{\ph}.\]
To compute the inner products involving $\al_2$ we use the fact
$(S+S\,R^t)\ch=R$,
as in the derivation of (\ref{SRS}), to write
\[((S+RS)(1-\ch),\,\ch\ps)=(1-\ch,\,R\,\ps)=(1-\ch,\,P-\ps)=\cP_1-a_1,\]
\[((S+RS)(1-\ch),\,\dl_t)=(1-\ch,\,R_t)=\cR_1,\]
\[((S+RS)(1-\ch),\,\dl_{\iy})=(1-\ch,\,R_{\iy})=0.\]
Using these, we find that
\[(\al_2,\,\be_1)=\hf(\cP_1-a_1+a_1\,\vet),\quad
(\al_2,\,\be_2)=\hf(\cR_1+a_1\,q\sep
-a_1\,c_{\ph}).\]

The GSE operator (\ref{4ops}) has the form (\ref{tenrep}) with $n=3$. But
notice that
\[((S+RS)\ve_t,\,\dl_{\iy})=(\ve_t,\,R_{\iy})=0,\quad
((S+RS)1,\,\dl_{\iy})=(1,\,R_{\iy})=0,\]
and $(Q\sep,\,\dl_{\iy})=c_{\ph}=0$ because for GSE $N$ is odd and so $\ph$ is
an odd
function. Thus the contribution to the determinant of the last term
in (\ref{4ops}) is 0 and we may discard it. The resulting operator is of the
form
(\ref{tenrep}) with $n=2$ and
\[\al_1=Q\sep,\quad \al_2=-\hf[(S+RS)\ve_t+a_4\,Q\sep],\quad\be_1=\ch\ps,
\quad\be_2=\dl_t,\]
where we have set
\[a_4:=(\ps,\,\ve_t).\]
The inner products are computed as for GOE above and we find that now
\[(\al_1,\,\be_1)=\vet,\quad(\al_1,\,\be_2)=q\sep,\]
\[(\al_2,\,\be_1)=-\hf(\cP_4-a_4+a_4\,\vet),\quad(\al_2,\,\be_2)=
-\hf(\cR_4+a_4\,q\sep).\]

Thus the determinants of the operators (\ref{1ops}) and (\ref{4ops}) are
expressible
in terms of the constants $a_1, a_4$ and $c_{\ph}$ and the as yet to be
determined
quantities $\vet,\,q\sep,\,\cP_1,\,\cR_1,\,\cP_4$, and $\cR_4$. The
determinants are
given by
\bq (1-\vet)\,(1-\hf\cR_1)-\hf( q\sep-c_{\ph})\,\cP_1,\label{p1ratio}\eq
\bq (1-\vet)\,(1+\hf\cR_4)+\hf q\sep\,\cP_4,\label{p4ratio}\eq
respectively. (The constants $a_1$ and $a_4$ drop out, as we see.)

\medskip\noi {\it The first set of differential equations:}\medskip

The derivation will not be quite self-contained because we shall refer
to \cite{FD}
for some results derived there.
First we have the analogues of (2.15)--(2.18) of \cite{FD},
\bq u\sep'=-q\,q\sep,\quad v\sep'=-q\,p\sep,\quad \vet'=-p\,q\sep,\quad w\sep'=
-p\,p\sep,\label{uveqs}\eq
which are proved in exactly the same way. (The primes denote $d/dt$). To derive
formulas for $q\sep'$ and $p\sep'$ we use the
displayed formula between (2.27) and (2.28) of \cite{FD}, which in this case
gives
\bq\left({\pl\ov\pl t}+{\pl\ov\pl x}+{\pl\ov\pl y}\right)\rh(x,y)=
-Q(x)\cdot(\ch P)(y)-P(x)\cdot(\ch Q)(y),\label{Dcom}\eq
where $\rh(x,y)$ is the kernel of $(I-S\ch)^{-1}$, in other words
$\dl(x-y)+R(x,y)$.
It follows from the above that
\[q\sep'={d\ov dt}\int\rh(t,y)\,\eph(y)\,dy=-\int{\pl\ov\pl
y}\rh(t,y)\,\eph(y)\,dy
-q\,(\ch P,\ve\ph)-p\,(\ch Q,\ve\ph).\]
The first term on the right side equals
\[\int\rh(t,y)\,\ph(y)\,dy=q.\]
We treat $p\sep$ similarly, and we find that we have derived the equations.
\bq q\sep'=q-q\,\vet-p\,u\sep,\quad p\sep'=p-q\,w\sep-p\,v\sep.\label{qpeqs}\eq

For the boundary conditions at $t=\iy$, observe that the four functions
$u\sep,\,
v\sep,\,\vet$ and $w\sep$ all vanish there, whereas
\[q\sep(\iy)=c_{\ph},\quad p\sep(\iy)=c_{\ps}.\]
One of these always vanishes, the first if $N$ is odd and the second if $N$ is
even.

It is easy to derive a first integral for our system of equations. Using the
equations
we find that
\[(p\sep\,q\sep)'=p\sep q(1-\vet)-p\sep pu\sep+q\sep(1-v\sep)-q\sep qw\sep\]
\[=v\sep'(1-\vet)-w\sep'u\sep+\vet(1-v\sep)-u\sep'w\sep
=-((1-v\sep)(1-\vet))'-(u\sep w\sep)',\]
and so, since all quantities vanish at $\iy$,
\[p\sep\,q\sep=1-(1-v\sep)(1-\vet)-u\sep w\sep.\]

\medskip\noi {\it The second set of equations:}\medskip

It follows from (\ref{Dcom}) that for $y\in J$
\[\left({\pl\ov\pl t}+{\pl\ov\pl x}+{\pl\ov\pl y}\right)R(x,y)=
-Q(x)\,P(y)-P(x)\, Q(y),\]
and so
\bq{\pl\ov\pl t}R(x,t)=-{\pl\ov\pl x}R(x,t)-p\,Q(x)-q\,P(x),\label{Rsder}\eq
and from this we obtain
\[\cR_1'=R(t,t)+\int_{-\iy}^t[-{\pl\ov\pl x}R(x,t)-p\,Q(x)-q\,P(x)]\,dx.\]
This gives our first equation,
\bq\cR_1'=-p\,\cQ_1-q\,\cP_1.\label{R'}\eq
By (2.10) of \cite{FD} we have
\[ {\pl Q\ov\pl t}=-R(x,t)\,q,\]
and so
\[\cQ_1'=q-q\,\int_{-\iy}^t R(x,t)\,dx.\]
We treat $\cP_1$ similarly, and so we have our other two equations,
\bq\cQ_1'=q\,(1-\cR_1),\quad\cP_1'=p\,(1-\cR_1).\label{QP'}\eq
At  $t=\iy$ our functions have the values
\bq\cR_1(\iy)=0,\quad\cQ_1(\iy)=
2\,c_{\ph},\quad\cP_1(\iy)=2\,c_{\ps}.\label{RQPiy}\eq

The system (\ref{R'}), (\ref{QP'}) has a first integral. If we multiply the
first equation by $\cR_1$ and use the second we obtain
\[\cR_1\cR_1'=(\cP_1'-p)\cQ_1+(\cQ_1'-q)\cP_1=(\cQ_1\cP_1)'+\cR_1'.\]
Integrating gives
\bq\hf\cR_1^2=\cQ_1\cP_1+\cR_1\,.\label{RQP}\eq
(Again the constant of integration is 0 because either $c_{\ph}$ or $c_{\ps}$
is 0.)
Differentiating (\ref{R'}) and using (\ref{QP'}) again give
\bq\cR_1''=-p'\,\cQ_1-q'\,\cP_1+2pq(\cR_1-1).\label{R''}\eq
Solving equations (\ref{R'}) and (\ref{R''}) for $\cQ_1$ and $\cP_1$ and
substituting
the results into (\ref{RQP}), we obtain a second-order differential equation
for
$\cR_1$:
\[-(\hf\cR_1^2-\cR_1)\,(q'p-p'q)^2=(p\,\cR_1''-p'\,\cR_1'+2p^2q\,(\cR_1-1))\,
(q\,\cR_1''-q'\,\cR_1'+2q^2p\,(\cR_1-1)).\]
To find a differential equation for $\cP_1$ we use the second relation of
(\ref{QP'}) to
express $\cR_1$ in terms of $\cP_1$, then use (\ref{R'}) to express $\cQ_1$ in
terms of
$\cP_1$, and then substitute these results into (\ref{RQP}). What results is
the equation
\[\cP_1\,\Bigl(({\cP_1'\ov p})'-q\cP_1\Bigr)={p\ov2}\,\Bigl(({\cP_1'\ov
p})^2-1\Bigr).\]

\medskip\noi {\it The third set of equations:}\medskip

To obtain the equations for $\cR_4,\,\cQ_4$ and $\cP_4$
we proceed almost exactly as in the last section, replacing the domain of
integration
$(-\iy,t)$ by $(-\iy,\iy)$, and inserting the factor $\ve(x-t)$
in the integrands. The only difference is that when we differentiate the
integrals we apply the product formula to the integrands. What results is the
system
\bq\cR_4'=-p\,\cQ_4-q\,\cP_4,\quad\cQ_4'=
-q\,(\cR_4+1),\quad\cP_4'=-p\,(\cR_4+1).
\label{R4eqs}\eq
The solutions of this system are obtained from the solutions of the last by
simply changing their signs. But notice that the values at $t=\iy$ are now
given by
\[\cR_4(\iy)=0,\quad\cQ_4(\iy)=-c_{\ph},\quad\cP_4(\iy)=-c_{\ps}.\]

\begin{center}{\bf VII. Scaling GOE and GSE at the Edge}\end{center}

The goal of this section is the computation of the limiting probability
distribution functions for the largest eigenvalue in the finite $N$ Gaussian
orthogonal and symplectic ensembles. The  probability distribution
functions for finite $N$ GOE, GUE and GSE are precisely the functions
$E_{\be}(0,\,(t,\iy))$ of Sec.~VI  with $\be=1,\,2$ and $4$, respectively. We
denote
them by $F_{N\be}(t)$, and limits we are interested in are
\[F_{\be}(s):=\lim_{N\ra\iy}F_{N\be}(\sqrt{2N}+{s\ov\sqrt2 N^{1/6}}).\]
In \cite{Airy} we showed that the limit exists when $\be=2$ and is given by
\bq F_2(s)=\exp\left(-\int_s^{\iy}(x-s)\,q(x)^2\,dx\right),\label{2pdf}\eq
where $q$ is the $P_{II}$ function determined by the differential equation
\[q''=s\,q+2\,q^3\]
 together with the condition $q(s)\sim {\rm Ai}(s)$ as $s\ra\iy$.
(For more details on this solution see \cite{Clarkson,Deift,Airy}.)\ \
 We shall show here that the limits exist for $\be=1$ and $4$ also, and that
\bq F_1(s)^2=F_2(s)\,e^{-\int_s^{\iy}q(x)\,dx},\label{1pdf}\eq
\bq
F_4(s/\sqrt{2})^2=F_2(s)\,
\lp{e^{\hf\int_s^{\iy}q(x)\,dx}+e^{-\hf\int_s^{\iy}q(x)\,dx}
\ov2}\rp^2.\label{4pdf}\eq
(Note the similarity to formulas (\ref{1ratio}) and (\ref{4ratio}).)

The reader must see our notational difficulty: $q$ denotes both a Painlev\'e
function and the function defined by (\ref{qetc}) of the last section. We
resolve
this difficulty by denoting the latter function here by $q_N$ (and denoting the
function $p$ of the last section by $p_N$), while retaining the notation $q$
for the
Painlev\'e II function.

We denote our scaling transformation by $\ta$, so that
\[\ta(x):=\sqrt{2N}+{x\ov\sqrt2 N^{1/6}}.\]
We think of $s$ as fixed, and the functions $q_N,\,p_N,
q\sep,\,u\sep,\cdots,\cP_4$ of
the last section as being associated with $t=\ta(s)$. We shall show that these
functions
of $s$ (perhaps after normalization) tend to limits as $N\ra\iy$ and that these
limits
satisfy systems of differential equations which are solvable in terms of the
$P_{II}$
function $q$. Substituting the values of these limits into (\ref{p1ratio}) and
(\ref{p4ratio}) will give (\ref{1pdf}) and (\ref{4pdf}). Everything will be a
consequence of the following:\medskip

(i) $\lim_{N\ra\iy}N^{-1/6}q_N=\lim_{N\ra\iy}N^{-1/6}p_N=q$.

(ii) The limits $\lim_{N\ra\iy}u\sep$ and $\lim_{N\ra\iy}\vet$ exist and are
equal.

(iii) The limits of $q\sep$ and $\cR_1,\cdots,\cP_4$ all exist. The limits of
$\cQ_1$
and $\cP_1$ differ by a constant as do the limits of $\cQ_4$ and $\cP_4$.

(iv) All of the above limits hold uniformly for bounded $s$.\medskip

These will be established below, but suppose for the moment that they are true.
Denote
the common limit in (ii) by $\bar u$, the limit of $q\sep$ by $\bar q$ and the
limits
of $\cR_1,\cdots,\cP_4$ by $\bar\cR_1,\cdots,\bar\cP_4$.

Let us rewrite the first
equations of (\ref{uveqs}) and (\ref{qpeqs}) using our present
notation and letting the prime now denote ${d\ov ds}\;(={d\ov dt}/\sqrt2
N^{1/6})$:
\[u\sep'=-{q_N\ov\sqrt2 N^{1/6}}\,q\sep,\quad q\sep'={1\ov\sqrt2 N^{1/6}}\,
(q_N-q_N\,\vet-p_N\,u\sep).\]
Taking the limits as $N\ra\iy$, using (i) and (ii), gives the system
\[\bar u'=-{q\ov\sqrt2}\,\bar q,\quad\bar q'={q\ov\sqrt2}\,(1-2\,\bar u).\]
(The interchange of the limit and the derivative is justified by the uniformity
of
convergence of these derivatives.) We remind the reader that when we apply our
finite
$N$ results to GOE and GSE we restrict $N$ to even or odd values, respectively,
and this
affects the boundary condition at $s=\iy$. In fact, we have
\[\bar u(\iy)=0,\quad \bar q(\iy)=\left\{\begin{array}{ll}1/\sqrt2&\mbox{in
GOE,}\\
&\\0&\mbox{in GSE.}\end{array}\right.\]
The reason for the first is that $u\sep$ vanishes at $\iy$ and
the reason for the second is that $q\sep(\iy)=c_{\ph}$, which vanishes when $N$
is odd
and can be shown to have the limit $1/\sqrt2$ as $N\ra\iy$ through even values.
Introduction of
\[\mu:=\int_s^{\iy}q(x)\,dx\]
as a new independent variable reduces our system of equations for $\bar u$ and
$\bar q$
to one with constant coefficients which is easily solved. We find that
\[\bar u=\hf(1-e^{-\mu}),\quad \bar q={1\ov\sqrt2}\, e^{-\mu}\quad \mbox{in
GOE,}\]
\[\bar u=\hf(1-\hf e^{\mu}-\hf e^{-\mu}),\quad \bar q=
{1\ov2\sqrt2}\, (e^{-\mu}-e^{\mu})\quad \mbox{in GSE.}\]

Next, we consider the limiting quantities $\bar\cR_1$, $\bar\cQ_1$ and
$\bar\cP_1$.
Recalling that these arise in GOE, when $N$ is even, we find from (\ref{RQPiy})
that
\[\bar\cR_1(\iy)=0,\quad\bar\cQ_1(\iy)=\sqrt2,\quad \bar\cP_1(\iy)=0,\]
and so, since by (iii) the last two functions differ by a constant,
\[\bar\cQ_1=\bar\cP_1+\sqrt2.\]
We find now that the limiting form of the system (\ref{R'}) and (\ref{QP'}) is
\[\bar\cR_1'=-{q\ov\sqrt2}(2\bar\cP_1+\sqrt2),\quad\bar\cP_1'=-{q\ov\sqrt2}
(\bar\cR_1-1).\]
The same substitution reduces this to a system with constant coefficients, and
we
find the solution to be
\[\bar\cR_1=1-e^{-\mu},\quad \bar\cP_1={1\ov\sqrt2}(e^{-\mu}-1).\]

Similarly, for $\be=4$ when $N$ is odd we find that
\[\bar\cR_4(\iy)=0,\quad\bar\cQ_4(\iy)=0,\quad \bar\cP_4(\iy)=-{\sqrt2\ov2},\]
that $\cQ_4=\cP_4+\sqrt2/2$, that the system is
\[\bar\cR_4'=-{q\ov\sqrt2}\,(2\bar\cP_4+{\sqrt2\ov2}),
\quad\bar\cP_4'=-{q\ov\sqrt2}
(\bar\cR_4+1),\]
and that the solution is
\[\bar\cR_4={1\ov4}e^{\mu}+{3\ov4}e^{-\mu}-1,\quad
\bar\cP_4={1\ov\sqrt2}\,({1\ov4}e^{\mu}-{3\ov4}e^{-\mu}-\hf).\]

If we substitute the limiting values we have found into the formulas
(\ref{p1ratio})
and (\ref{p4ratio}) which give the values of the ratios $F_{N1}(t)^2/F_{N2}(t)$
and
$F_{N4}(t/\sqrt{2})^2/F_{N2}(t)$ we obtain formulas (\ref{1pdf}) and
(\ref{4pdf}).

It remains to establish our claims (i)--(iv) above. We indicates by a subscript
$\ta$ the result of scaling either a function or an operator. Thus,
\[S_{\ta}:=\ta\circ S\circ\ta,\quad \ph_{\ta}:=\ph\circ\ta,\quad{\rm etc..}\]
It follows from results on the asymptotics of Hermite polynomials that
\bq\lim_{N\ra\iy}N^{-1/6}\ph_{\ta}(x)=\lim_{N\ra\iy}N^{-1/6}\ps_{\ta}(x)=A(x)
\label{phiscale}\eq
uniformly for bounded $s$, where $A(x)$ denotes the Airy function Ai($x$), and
that there are estimates
\bq N^{-1/6}\ph_{\ta}(x)=O(e^{-x}),\quad N^{-1/6}\ps_{\ta}(x)=O(e^{-x})\label
{phiest}\eq
which hold uniformly in $N$ and for $x$ bounded below. (There is a better
bound, in
which $x^{3/2}$ appears in the exponent rather than $x$, but this one is more
than
good enough for our purposes. See \cite{Ol}, p. 403.)

To obtain the scaling limits of $S$ and other quantities we shall make use of
the identity
\bq S(x,y)=\izy [\ph(x+z)\,\ps(y+z)+\ps(x+z)\,\ph(y+z)]\,dz\label{Nkernrep}\eq
analogous to formula (4.5) of \cite{Airy},
\[{A(x)\,A'(y)-A'(x)\,A(y)\ov x-y}=\izy A(x+z)\,A(y+z)\,dz\]
and proved in the same way: The formula on the top of p. 43 of \cite{FD} gives
\[\lp{\pl\ov\pl x}+{\pl\ov\pl y}\rp S(x,y)=-\ph(x)\,\ps(y)-\ps(x)\,\ph(y),\]
and the same operator applied to the right side of (\ref{Nkernrep}) equals
\[\izy \lp{\pl\ov\pl x}+{\pl\ov\pl
y}\rp[\ph(x+z)\,\ps(y+z)+\ps(x+z)\,\ph(y+z)]\,dz\]
\[=\izy{\pl\ov\pl
z}[\ph(x+z)\,\ps(y+z)+\ps(x+z)\,\ph(y+z)]\,dz=-\ph(x)\,\ps(y)-
\ps(x)\,\ph(y).\]
Hence the two sides of (\ref{Nkernrep}) differ by a function of $x-y$ and this
function
must vanish since both sides tend to 0 as $x$ and $y$ tend to $\iy$
independently.

If an operator $L$ has kernel $L(x,y)$ then $L_{\ta}$ has kernel
\[{1\ov\sqrt2 N^{1/6}}L(\ta(x),\ta(y)),\]
and so from (\ref{Nkernrep}) we see that the kernel of $S\st$ has the
representation
\[S_{\ta}(x,y)={1\ov\sqrt2 N^{1/6}}\izy [\ph(\ta(x)+z)\,\ps(\ta(y)+z)+
\ps(\ta(x)+z)\,\ph(\ta(y)+z)]\,dz,\]
and the substitution $z\ra z/\sqrt2 N^{1/6}$ gives
\bq S_{\ta}(x,y)={1\ov 2 N^{1/3}}\izy [\ph\st(x+z)\,\ps\st(y+z)+
\ps\st(x+z)\,\ph\st(y+z)]\,dz.\label{Sta}\eq
The asymptotic formulas (\ref{phiscale}), and the estimates (\ref{phiest}),
show that
\[\lim_{N\ra\iy}S_{\ta}(x,y)=\izy A(x+z)\,A(y+z)\,dz\]
pointwise, and in various function space norms as well. This will be very
useful.
The right side, we know, is the Airy kernel.
We have been using a bar as notation for a limit, so we write the above as
\bq S_{\ta}(x,y)\ra \bar S(x,y)=\izy A(x+z)\,A(y+z)\,dz.\label{kernlim}\eq

The fact that this converges in the space $L_2(s,\iy)\tn L_2(s,\iy)$ implies
that
the operator $\ch_{\ta}S_{\ta}\ch_{\ta}$ converges in the norm of
operators on $L_2(\bR)$ to the operator $\bch\bar S\bch$, where $\bch=
\ch_{(s,\iy)}$. Since $I-\bch\bar S\bch$ is invertible it follows that also
\bq(I-\ch_{\ta}S_{\ta}\ch_{\ta})\inv\ra (I-\bch\bar S\bch)\inv\label{conv}\eq
in operator norm. The same is true if the space $L_2(\bR)$ is replaced by
$L_1(\bR)$,
as is seen if we use the fact that the norm of an operator $\bch L\bch$
on this space is at most
\[\int_s^{\iy} \sup_{y\geq s}|L(x,y)|\,dx.\]
and that (\ref{kernlim}) holds with this function space norm as well.

Let us compute the scaling limit of $q_N$. We have
\[q_N=(I-S\ch)\inv\ph\,(t)=\ph(t)+S\ch(I-\ch S\ch)\inv\ph\,(t)=
\ph_{\ta}(s)+S_{\ta}\ch_{\ta}(I-\ch_{\ta}S_{\ta}\ch_{\ta})\inv\ph_{\ta}\,(s).\]
It follows from (\ref{phiscale}) and (\ref{phiest}) that $N^{-1/6}\ph_{\ta}
\ra A$ pointwise and in $L_2(s,\iy)$. The second of these facts, together with
(\ref{conv}), implies that
\[(I-\ch_{\ta}S_{\ta}\ch_{\ta})\inv N^{-1/6}\ph_{\ta}\ra
(I-\bch\bar S\bch)\inv A\]
in $L_2$, and then using the fact that (\ref{kernlim}) holds in $L_2(s,\iy)$
for
fixed $x$, we deduce that
\[S_{\ta}\ch_{\ta}(I-\ch_{\ta}S_{\ta}\ch_{\ta})\inv N^{-1/6}\ph_{\ta}\ra
\bar S\bch(I-\bch\bar S\bch)\inv A\]
pointwise. Putting these together shows that
\[\lim_{N\ra\iy}N^{-1/6}q_N=A(s)+\bar S\bch(I-\bch\bar S\bch)\inv A\,(s)
=(I-\bch\bar S\bch)\inv A\,(s).\]
The right side was precisely the {\em definition} of $q$ (it transpired that it
was a
Painlev\'e function), so we have shown that
\[\lim_{N\ra\iy}N^{-1/6}q_N=q.\]
Note that since by (\ref{phiscale}) both $\ph$ and $\ps$ have the same scaling
limit,
the above argument applied to $p_N$ leads to the same result,
\[\lim_{N\ra\iy}N^{-1/6}p_N=q.\]
This gives assertion (i). The uniformity assertion in (iv), for these and the
limits established below, we leave to the reader.

Next we consider $u\sep$ and $\vet$. Recalling the definition
(\ref{cdef}) we write
\[\ve\ph\,(x)=c_{\ph}-\int_x^{\iy}\ph(y)\,dy,\]
and so
\[(\ve\ph)_{\ta}(x)=c_{\ph}-\int_{\ta(x)}^{\iy}\ph(y)\,dy=c_{\ph}-{1\ov\sqrt2
N^{1/6}}
\int_x^{\iy}\ph_\tau(y)\,dy.\]
As $N\ra\iy$ the constant $c_{\ph}$ converges (in fact to $1/\sqrt2$) and the
second
term converges in $L_2(s,\iy)$ (in fact to $-\int_x^{\iy}A(y)\,dy/\sqrt2$).
Thus the
function $(\ve\ph)_{\ta}$ converges in the space $\bR+L_2(s,\iy)$.
Also,
\bq N^{-1/6}Q_{\ta}=N^{-1/6}(I-\ch\st S\st\ch\st)\inv\ph_{\ta}\ra
(I-\bch\bar S\bch)\inv A\label{Qlim}\eq
in $L_1(s,\iy)\cap L_2(s,\iy)$. It follows from these limit relations that
\[u\sep=(Q,\,\ch\ve\ph)={1\ov\sqrt2 N^{1/6}}(Q_{\ta},\,(\ch\ve\ph)_{\ta})\]
converges as $N\ra\iy$. Notice also that since $\ph$ and $\ps$ have the same
scaling
limit, (\ref{Qlim}) holds with $Q$ replaced by $P$ on the left side, from which
it
follows that $\vet$ has the same limiting value as $u\sep$. This establishes
(ii).

Finally we come to the quantities $\cR_1,\cdots,\cP_4$. These are trickier
since,
although our functions scale nicely on $(s,\iy)$ for fixed $s$, they do not
scale
uniformly on $(-\iy,\,\iy)$. In a sense we have to separate out the parts that
get
integrated over $(-\iy,\,\iy)$. Beginning with $\cR_1$, we use
\[R(x,t)=(I-S\ch)\inv S(x,t)=S\ch(I-\ch S\ch)\inv S(x,t)+S(x,t),\]
and write our integrals over $(-\iy,\,t)$ as integrals over $(-\iy,\,\iy)$
minus
integrals over $(t,\,\iy)$. Thus
\[\cR_1=\iyy S\ch(I-\ch S\ch)\inv S(x,t)\,dx-\int_t^{\iy}S\ch(I-\ch S\ch)\inv
S(x,t)\,dx\]
\[+\iyy S(x,t)\,dx-\int_t^{\iy}S(x,t)\,dx\]
\[=\iyy S\st\ch\st(I-\ch\st S\st\ch\st)\inv S\st(x,s)\,dx
-\int_s^{\iy}S\st\ch\st(I-\ch\st S\st\ch\st)\inv S\st(x,s)\,dx\]
\bq +\iyy S\st(x,s)\,dx-\int_s^{\iy}S\st(x,s)\,dx.\label{cR1}\eq
(The factors $1/\sqrt2 N^{1/6}$ arising from the variable change are
incorporated in the expression for $S\st(x,s)$.)
We think of the first integral on the right side of (\ref{cR1}) as the inner
product
of the functions
\bq\iyy S\st(x,\cdot)\,dx\quad\mbox{and}\quad(I-\ch S\st\ch\st)\inv
S\st(\cdot,s).\label{ip}\eq
on $(s,\iy)$. Now it follows from (\ref{Sta}) that
\bq\iyy S\st(x,y)\,dx={1\ov\sqrt2 N^{1/6}}\lp
c_{\ph}\int_y^{\iy}\ps\st(z)\,dz+
c_{\ps}\int_y^{\iy}\ph\st(z)\,dz\rp,\label{Sint}\eq
which converges in $L_2(s,\iy)$ as $N\ra\iy$. Also $S\st(\cdot,s)$ converges in
$L_2(s,\iy)$, so the same is true of the second function in
(\ref{ip}). Thus the inner product itself converges, and this shows that the
first
term on the right side of (\ref{cR1}) converges. The second term (\ref{cR1}) is
treated
in much the same way---the only difference is that the first integral in
(\ref{ip})
is taken over $(s,\iy)$. The next term on the right side of (\ref{cR1}) is
simple---to show that it converges requires only the representation
(\ref{Sint})
 with $y=s$---and the last term is analogous to this one.

Turning to $\cQ_1$, we write it similarly as
\[\iyy S\ch(I-\ch S\ch)\inv\ph(x)\,dx-\int_t^{\iy}S\ch(I-\ch S\ch)\inv
\ph(x)\,dx
+\iyy\ph(x)\,dx-\int_t^{\iy}\ph(x)\,dx\]
\[=\iyy S\st\ch\st(I-\ch\st S\st\ch\st)\inv{\ph\st(x)\ov\sqrt2 N^{1/6}}\,dx
-\int_s^{\iy}S\st\ch\st(I-\ch\st S\st\ch\st)\inv{\ph\st(x)\ov\sqrt2
N^{1/6}}\,dx\]
\[+2\,c_{\ph}-\int_s^{\iy}{\ph\st(x)\ov\sqrt2 N^{1/6}}\,dx.\]
We treat the integrals just as we did those for $\cR_1$, with $\ph\st/\sqrt2
N^{1/6}$
replacing $S\st(\cdot,s)$, to show that they have limits as $N\ra\iy$ and, of
course,
$c_{\ph}$ also has a limit. The quantity $\cP_1$ is handled similarly, with
$\ps$ replacing $\ph$
evrywhere. That the limits of $\cP_1$ and $\cQ_1$ differ by a constant follows
immediately from the fact that $\ps$ and $\ph$ have the same scaling limit on
$(s,\,\iy)$. The discussion of the quantities $\cR_4,\,\cQ_4$ and $\cP_4$ is
entirely
analogous, and (iii) is established.

\begin{center} {\bf VIII. Justification for the Determinant Manipulations}
\end{center}

In this final section we give the rigorous justification for the determinant
manipulations in Sections II and III. We begin with GSE, which is slightly
easier.
The quantity of interest is the determinant for the operator on $L_2(J)$ with
kernel
\[\hf \ba S(x-y)&DS(x-y)\\&\\IS(x-y)&S(x-y)\ea,\]
where
\[S(x)={\sin x\ov\pi x},\quad DS(x)=S'(x),\quad IS(x)=\int_0^xS(y)\,dy.\]
This kernel is smooth and so the operator is trace class. Our determinant is
the same
as that for the operator on $L_2({\bf R})$
\bq\hf\ba \ch S\ch&\ch DS\ch\\&\\\ch IS\ch&\ch S\ch\ea,\label{gseop1}\eq
where $S, DS, IS$ are the operators with corresponding difference kernels and
$\ch$
is multiplication by $\ch_J$. We shall show that we can remove all the
operators $\ch$
which appear on the left, if we interpret $S, DS$, and $IS$ as acting between
appropriate spaces.

Recall that the Sobolev space $H_1$ is given by
\[\{f\in L_2: f\mbox{ is absolutely continuous and }f'\in L_2\},\]
with
\[\norm f\norm_{H_1}=\sqrt{\norm f\norm_{L_2}^2+\norm f'\norm_{L_2}^2}.\]
The mapping
\bq f\ra f+D\,f \qquad(D={d\ov dx})\label{isometry}\eq
is an isometry from $H_1$ onto $L_2$.\medskip

\noi{\it Lemma 1}. $S\ch$ and $DS\ch$ are trace class operators from $L_2$ to
$H_1$.
\medskip

\noi{\it Proof}. If the kernel $K(x,y)$ of an operator $K$ satisfies
\[\int_{\bf R}\int_J\,|K(x,y)|^2\,dy\,dx<\iy,\quad
\int_{\bf R}\int_J\,|{\pl\ov\pl y} K(x,y)|^2\,dy\,dx<\iy,\]
then $K\ch_J$ is trace class for any set $J$ of finite measure. (See, for
example,
pp.~118--119 of \cite{DS}.)\ \ It is
immediate that $D^nS\ch:L_2\ra L_2$ is trace class for each $n\geq0$. If we
recall the
isometry (\ref{isometry}) we see that to show that $D^nS\ch:L_2\ra H_1$ is
trace
class it is enough to show that $(I+D)D^nS\ch:L_2\ra L_2$ is. But we know
this.\medskip

Next we enlarge our spaces $L_2$ and $H_1$ by adjoining a single function to
each:
\[\Ltl:=\{f+c\,\ve:f\in L_2,\,c\in{\bf C}\},\quad\Htl:=\{f+c\,IS:f\in
H_1,\,c\in{\bf C}\},\]
which are Hilbert spaces when endowed with the norms
\[\sqrt{\norm f\norm_{L_2}^2+|c|^2},\quad\sqrt{\norm f\norm_{H_1}^2+|c|^2},\]
respectively.\medskip

\noi{\it Lemma 2}. $IS\ch$ is a trace class operator from $L_2$ to $\Htl$.
\medskip

\noi{\it Proof}. Integration by parts on the constituent intervals of $J$ shows
that for all $f\in L_2$
\bq(IS)(f)(x)=\sum\,(-1)^k\,IS(x-a_k)\,(If)(a_k)+
\int\,S(x-y)\,(If)(y)\,\ch_j(y)\,dy,
\label{IS}\eq
where $(If)(x):=\int_0^x\,f(y)\,dy$. The map $f\ra(If)\cdot\ch_J$ is bounded
from $L_2$ to $L_2$ and so by Lemma 1 the integral on the right represents a
trace
class operator from $L_2$ to $H_1$. The maps $f\ra(If)(a_i)$ are continuous
linear
functionals on $L_2$ and each function $IS(x-a_i)$ belongs to $\Htl$ since
\[IS(x-a_i)-IS(x)\in H_1.\]
So the sum on the right side represents a finite rank, and hence trace class,
operator
from $L_2$ to $\Htl$. This completes the proof.\medskip

Let us consider the operator represented by the matrix
\bq\hf\ba S\ch&DS\ch\\&\\IS\ch&S\ch\ea,\label{gseop2}\eq
which is the same as (\ref{gseop1}) except that all factors $\ch$ that appeared
on
the left were removed. Since $\ch:\Ltl\ra L_2$ is bounded, it follows from
Lemmas 1
and 2 that this is a trace class operator from $L_2\ds\Ltl$ to itself.
(Actually, of
course, the operator is trace class from $L_2\ds\Ltl$ to $H_1\ds\Htl$,
but we don't use this.)

We use now, and several times below, the fact that the determinant for an
operator
product $AB$ is the same as for $BA$ as long as one of the two factors is trace
class and the other is bounded. They do {\em not} have to act on the same
Hilbert
space; one operator can map a space $H$ to a space $H'$ as long as the other
maps
$H'$ to $H$. The two products then act on the different spaces $H$ and
$H'$.\medskip

\noi{\it Lemma 3}. The determinants for the operator (\ref{gseop1}) on $L_2\ds
L_2$
and the operator (\ref{gseop2}) on $L_2\ds\Ltl$ are equal.\medskip

\noi{\it Proof}. Because $\ch$ is idempotent we can insert a factor
$\ba\ch&0\\0&\ch\ea$ on the right side of (\ref{gseop2}) without changing it.
We can
bring this factor  aound to the other side and deduce that the determinant for
(\ref{gseop2}) as an operator on $L_2\ds\Ltl$ is the same as that for
(\ref{gseop1})
as an operator on $L_2\ds\Ltl$. But the range of this operator is contained in
$L_2\ds L_2$, so the determinant for it is the same when considered an operator
on this space.\medskip

It follows from the lemma that we can replace (\ref{gseop1}) by (\ref{gseop2}).
Since
$D(IS)=S$ we can write it as the product
\[\hf\ba D&0\\&\\0&I\ea\ba IS\ch&S\ch\\&\\IS\ch&S\ch\ea.\]
It follows from Lemmas 1 and 2 that the factor on the right is trace class from
$L_2\ds\Ltl$ to $\Htl\ds\Htl$ while the factor on the left is bounded from
$\Htl\ds\Htl$ to $L_2\ds\Ltl$ since $D:\Htl\ra L_2$ is bounded. It follows that
the
determinant for (\ref{gseop2}) on $L_2\ds\Ltl$ is the same as the determinant
for
\[\hf\ba IS\ch&S\ch\\&\\IS\ch&S\ch\ea\ba D&0\\&\\0&I\ea=
\hf\ba IS\ch D&S\ch\\&\\IS\ch D&S\ch\ea\]
on $\Htl\ds\Htl$. The matrix entries are operators on the same
space, $\Htl$, so that we can perform what amounts to row and column operations
on
them. (This was not an accident!) Multiplying
on the left by the matrix $\ba I&0\\-I&I\ea$ and on the right by its inverse,
$\ba I&0\\I&I\ea$, we see that the last determinant is the same as that for
\[\hf\ba IS\ch D+S\ch&S\ch\\&\\0&0\ea,\]
and so it is the determinant {\em of}
\[\ba I-\hf IS\ch D-\hf S\ch&-\hf S\ch\\&\\0&I\ea=
\ba I&-\hf S\ch\\&\\0&I\ea\ba I-\hf IS\ch D-\hf S\ch&0\\&\\0&I\ea.\]
Both factors on the right are operators of the form $I+$ trace class operator
on
$\Htl\ds\Htl$, and the factor on the left is of the form $I+$ nilpotent
operator and
so has determinant 1. Hence the determinant of the product equals the
determinant of
the second factor, which equals
\[I-\hf IS\ch D-\hf S\ch\]

We rewrite the operator as
\bq I-S\ch-\hf(IS\ch D-S\ch).\label{op1}\eq

A variant of (\ref{IS}) is
\[(IS\ch)(f')(x)=\sum\,(-1)^k\,IS(x-a_k)\,f(a_k)+
\int\,S(x-y)\,f(y)\,\ch_j(y)\,dy,\]
which gives
\[ IS\ch D-S\ch=\sum (IS)_k\tn\dl_k,\]
where $(IS)_k(x):=IS(x-a_k)$ and $\dl_k(x):=\dl(x-a_k)$. The tensor product
denotes
the operator which takes a function in  $\Htl$, evaluates it at $a_k$ and
multiplies
this by $(IS)_k$. A similar interpretation holds for any tensor product $u\tn
v$
where $u$ belongs to a Hilbert space and $v$ to its dual space.

We extend the domain of $S$ to all of $\Ltl$ by defining
\[(Sf)(x)=\int S(x-y)\,f(y)\,dy,\]
the integral being conditionally convergent. It is easy to see that
$S:\Ltl\ra\Htl$ and $(IS)_k=S\ve_k$, where $\ve_k(x):=\ve(x-a_k)$. Thus we can
write
(\ref{op1}) as
\bq I-S\ch-\hf\sum(-1)^k\,S\ve_k\tn\dl_k.\label{op2}\eq
Recall that our operators act on $\Htl$. Now $I-S\ch$ is invertible as an
operator
on this space as well as on $L_2$, since the eigenfunctions of $S\ch$
belonging to nonzero eigenvalues are the same for the two spaces. For the same
reason
both interpretations of the operator give the same value for the determinant.
We
denote the inverse of the operator, as before, by $I+R$. Factoring out $I-S\ch$
shows
that the determinant of (\ref{op2})
equals $E_2(0,J)$ times the determinant of
\bq I-\hf\sum(-1)^k\,(S+RS)\ve_k\tn\dl_k.\label{op3}\eq
Recall the definition $R_k(x):=R(x,a_k).$\medskip

\noi{\it Proposition}. The determinant of (\ref{op3}) on $\Htl$ equals the
determinant
of
\bq I-\hf\sum(-1)^k\,\ve_k\tn R_k\label{op4}\eq
on $\Ltl$.\medskip

\noi{\it Proof}. We use inner product notation $(u,v)$ to denote the action of
a dual
vector $v$ on a vector $u$. The determinants of (\ref{op3}) and (\ref{op4}) are
scalar determinants whose entries contain the inner products
\[((S+RS)\ve_j,\dl_k),\quad(\ve_j,R_k),\]
respectively, in position $(j,k)$. We shall show that these are equal.

We begin with the observation that $R^t\ch=\ch R$ when these are thought of as
acting
on $L_2$. This is so because $R$ is the resolvent operator for $S\ch$ and $S$
is
symmetric. It follows from this that if $f$ and $g$ belong to $L_2$, with $g$
supported in $J$, then
\[((S+RS)f,g)=(f,(S\ch+S\ch R)g)=(f,Rg),\]
the last by the resolvent identity. Suppose $h$ has integral 1 and is compactly
supported (in ${\bf R}^+$ if $k$ is odd, in ${\bf R}^-$ if $k$ is even), set
$g(x)=n\,h(n(x-a_k))$ and let $n\ra\iy$. We obtain
\[((S+RS)f,\dl_k)=(f,R_k).\]
Replace $f$ by $f_n:=\ve_j\ch_{[-n,n]}$ and let $n\ra\iy$. Since
\[(S+RS)f_n\ra(S+RS)\ve_j\]
uniformly on compact sets we deduce the desired identity
\[((S+RS)\ve_j,\dl_k)=(\ve_j,R_k).\]
Thus (\ref{E4JDet}) is completely proved.

We now turn to GOE. The reason this is slightly awkward is that the operator
$K$ on
$L_2(J)$ with kernel
\bq\ba S(x-y)&DS(x-y)\\&\\IS(x-y)-\ve(x-y)&S(x-y)\ea\label{goekernel}\eq
is not trace class . So its classical Fredholm
determinant, which is what we want, is not given by det$(I-K)$, which
is not defined. It is instead given in terms of its regularized determinant
$det_2$
by the formula
\[\mbox{det}_2(I-K)\;e^{-{\rm tr}\,K},\]
where $tr\,K$ denotes the sum of the integrals of the diagonal entries of the
kernel
of $K$. Rather than deal with regularized determinants we approximate the
kernel by
a smooth kernel, evaluate the resulting Fredholm determinant, and pass to the
limit
at the end. Thus we replace the term $\ve(x-y)$ in (\ref{goekernel}) by
$\eta_n(x):=\eta(n(x-y))$,
where $\eta$ is a smooth function which equals $\ve$ outside a finite interval.
Then
the resulting operator $K_n$ is trace class, and $\det_2(I-K)\,e^{-{\rm
tr}\,K}$ is
equal to the limit of $\det(I-K_n)$ as $n\ra\iy$.

Proceeding as before, we find that $\det(I-K_n)$ is equal to the determinant of
\[\ba I-IS\ch D-S\ch&-S\ch\\&\\\eta_n\ch D&I\ea\]
on $\Htl\ds\Htl$, where $\eta_n$ denotes convolution by $\eta_n(x)$. The
operator can
be factored as
\[\ba I&-S\ch\\&\\0&I\ea\ba I-IS\ch D-S\ch+S\ch\eta_n\ch D&0\\&\\0&I\ea
\ba I&0\\&\\\eta_n\ch D&I\ea,\]
all factors being of the form $I+$ trace class operator on $\Htl\ds\Htl$,
and so its determinant equals the determinant of the operator
\[I-S\ch-IS\ch D+S\ch\eta_n\ch D\]
on $\Htl$. But $\ch\eta_n\ch\ra\ch\ve\ch$ in the norm for operators on $L_2$
(where
$\ve$ denotes convolution by $\ve(x)$) and it follows from this and the second
part of
Lemma 1 that the above operator converges to
\bq I-S\ch-IS\ch D+S\ch\ve\ch D\label{op5}\eq
in the trace norm for operators on $\Htl$. Consequently it is the determinant
of this
which will be our final answer.

We have already seen (c.f. (\ref{op1}) and (\ref{op2})) that
\[IS\ch D=S\ch+\sum(-1)^k\,S\ve_k\tn\dl_k.\]
But
\[(\ve\ch D\,f)(x)=\int_J\ve(x-y)\,f'(y)\,dy=(\ch
f)(x)+\sum(-1)^k\,\ve_k(x)\,f(a_k),\]
and so
\[S\ch\ve\ch D=S\ch+\sum(-1)^k\,S\ch\ve_k\tn\dl_k.\]
Thus
\[IS\ch D-S\ch\ve\ch D=\sum(-1)^k\,S(1-\ch)\ve_k\tn\dl_k.\]
And now, just as at the end of GSE, we conclude that the determinant of
(\ref{op5})
equals $E_2(0,J)$ times the determinant of
\[I-\sum(-1)^k\,(S+RS)(1-\ch)\ve_k\tn\dl_k,\]
which in turn equals the determinant of
\[I-\sum(-1)^k\,(1-\ch)\ve_k\tn R_k.\]
This completes the justification of (\ref{E1JDet}).
\bigskip
\begin{center}{\bf Acknowledgements}\end{center}

The authors wish to thank Professor Freeman Dyson for some helpful comments
in the early stages of this work.  Part of this work was done at Imperial
College and the  first author  thanks
the Mathematics Department of Imperial College and in particular
Dr.~Yang Chen for  kind hospitality  and  the Science and Engineering Research
Council of UK for the award of a Visiting Fellowship that made this stay
possible.
This work was supported in part by
the National Science Foundation through grants DMS--9303413 and DMS--9216203.

\end{document}